\documentclass[mathleft
]{an}
\usepackage{graphicx}
\usepackage{times}
\usepackage{natbib}
\bibpunct{(}{)}{;}{a}{}{,}
\begin{document}

\Pagespan{789}{}
\Yearpublication{2006}%
\Yearsubmission{2005}%
\Month{11}%
\Volume{999}%
\Issue{88}%

\title{On the likelihood-ratio test applied in asteroseismology \\
for mode identification}

\author{D. Salabert\inst{1,2}\fnmsep\thanks{Corresponding author:
  \email{salabert@iac.es}\newline}
\and  R.~A. Garc\'ia\inst{3}
\and  S. Mathur\inst{4}
}
\titlerunning{On the likelihood-ratio test in asteroseismology}
\authorrunning{D. Salabert et al.}
\institute{Instituto de Astrof\'isica de Canarias,  E-38200 La Laguna, Tenerife, Spain
\and 
Departamento de Astrof\'isica, Universidad de La Laguna, E-38205 La Laguna, Tenerife, Spain
\and 
Laboratoire AIM, CEA/DSM-CNRS, Universit\'e Paris 7 Diderot, IRFU/SAp, Centre de Saclay, F-91191 Gif-sur-Yvette, France
\and
High Altitude Observatory, National Center for Atmospheric Research, P.O. Box 3000, Boulder, Colorado 80307-3000, USA}



\abstract{The identification of the solar-like oscillation modes, as measured by asteroseismology, is a necessary requirement in order to infer the physical properties of the interior of the stars. Difficulties occur when a large number of modes of oscillations with a low signal-to-noise ratio are observed. In those cases, it is of common use to apply a likelihood-ratio test to discriminate between the possible scenarios. We present here a statistical 
analysis of the likelihood-ratio test and discuss its accuracy to identify the correct modes. We use the AsteroFLAG artificial stars, simulated over a range of magnitude, inclination angle, and rotation rate. We show that the likelihood-ratio test is appropriate up to a certain magnitude (signal-to-noise ratio). }

\maketitle

\section{Introduction}
The recent development of space-based instruments such as the Convection, Rotation, and planetary Transits mission \citep[CoRoT;][]{Michel08} and the NASA's Kepler mission \citep{Borucki09} as well as the organized ground-based campaigns \citep{Arentoft08} are producing a huge volume of asteroseismic observations of unprecedented quality. Thus, it is now possible to correctly detect individual pressure (p) driven modes and to describe the oscillation modes with Lorentzian profiles, as it is commonly done for the Sun \citep[e.g.,][]{Chaplin06}. 
However, in the stellar case, there is a substantial difference: while in the Sun, the inclination of the rotation axis and the surface rotation rate are known, in most of the stars analyzed up to now, these two parameters are unknown. Moreover, as these two parameters are highly correlated \citep{Gizon03,Ballot06}, in order to improve the stability of the fits, the traditional pair-by-pair fitting methodology followed in the Sun is changed to a global strategy in which all the modes are fitted at the same time along with the inclination angle and one value for the rotational splitting. The first application of this strategy was introduced in asteroseismology by \citet{Appourchaux08} and it has  been since succesfully applied to the CoRoT solar-like targets in which the signal-to noise ratio was high enough \citep{Barban09, Garcia09, Deheuvels10}. Indeed, some of the measured CoRoT stars were too faint to perform such kind of analyses \citep{Mosser09,Mathur10a}. Moreover, the global fitting has also been applied to the first Kepler solar-like targets \citep{Chaplin10}, as well as the latest ground-based observational campaign of Procyon \citep{Bedding10b}.

Everything would be great if we were able to identify the individual modes (or mode-tagging) before performing the peak-bagging. Indeed, for most of the observations, a clear identification of the modes (or of the ridges) appears to be a difficult task and it is then necessary to perform the peak-fitting using the two possible identifications. The possible scenarios (or taggings) are discriminated by comparing a posteriori the likelihoods of the minimization, the highest likelihood being chosen as the correct mode identification. \citet{Benomar09} using a longer time series -- which means a better overall signal-to-noise ratio -- demonstrated by comparing the likelihoods that the first p-mode identification done by \citet{Appourchaux08} of the CoRoT target HD49933 was wrong. However, we present in this work that the so-called {\it likelihood-ratio test} has certain limits depending on both the magnitude of the star and the length of the observations (i.e., the overall signal-to-noise ratio of the modes in the power spectrum). It is important to remember that an incorrect mode tagging would have very bad consequences on the inferences of the stellar properties \citep[e.g.,][]{creevey07,Stello09} .

\section{Data and analysis}
We used simulated data generated within the AsteroFLAG team for hare-and-hounds exercises \citep{Chaplin08}. In this work, we concentrate on one of these stars: Pancho, whose fundamental parameters $\log(g)$, $\log(Z/X)$, and  T$\mathrm{_{eff}}$ are given in Table~\ref{table:1}. 
The following analysis is performed over a range of:
\begin{itemize}
\item apparent magnitude M$_v$: 9, 10, 11, and 12;
\item inclination angle: $30^{\circ}$ and $60^{\circ}$;
\item rotation rate: 1.5 (slow), 3.0 (medium), and 5.0 (high) $\mu$Hz.
\end{itemize} 
The original simulated ~$\sim$ 3-year time series (with a 60-second cadence) were divided into 3, 12, and 36 sub-series of 365, 93, and 31 days respectively. The global parameters $\Delta\nu$,  Freq(min), Freq(max), $\nu_{\mathrm{max}}$, and A$_{\mathrm{max}}$ derived from 31-day time series are given in Table~\ref{table:2} \citep{Mathur10b}. A Maximum-Likelihood Estimator (MLE) global fitting of the $l=0,1,2$ modes using the strategy defined in \citet{Appourchaux08} is performed on each power spectrum. The number of fitted modes is fixed as a function of the length $T$ of the sub-series as follows:
\begin{itemize}
\item 31 days: 6 and 8 large separations;
\item 93 days: 10 and 14 large separations;
\item 365 days: 14 large separations.
\end{itemize}

We define two different mode taggings: {\sc a} for the correct one, and {\sc b} for the incorrect one. Two different peak-fitting approaches were used as well: one with no Bayesian constraints, and one with three Bayesian constraints, defined as:
\begin{itemize}
\item inclination angle $\pm  15 ^{\circ}$ of the simulated value;
\item rotational splitting $\pm1.5 \mu$Hz of the simulated value;
\item Full-Width-at-Half-Maximum ({\sc fwhm}) of the high order modes $\geq 2 \mu$Hz  (to avoid fitting spikes instead of wider modes).
\end{itemize}

We have run our fitting code on the {\sc godunov} cluster at CEA/SAp. It has 15 bi-processors nodes and 15 dual core bi-processor nodes, for a total of 90 cores connected through a gigabit network and running IDL. We have used typically 15 to 30 cores for 2 to 5 days to run all the fittings. Indeed, we run in parallel several fits of the same time series using different strategies.

\section{Comparison with the input frequencies}
\begin{figure*}
\centering
\includegraphics[scale=0.43]{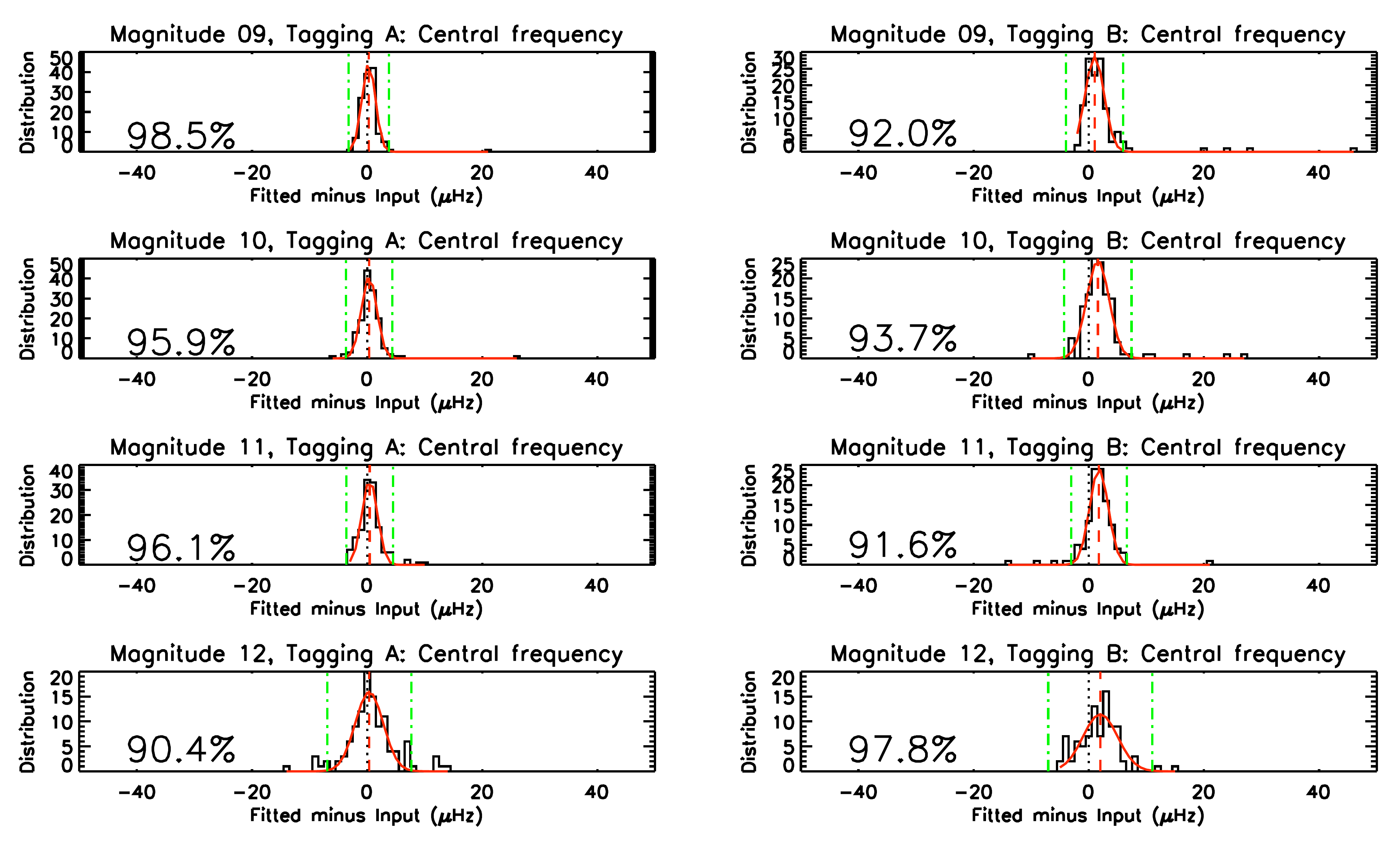}
\caption{Histograms of the frequency differences between {\sc fits} and {\sc inputs} for the $l = 0$ modes in the case of the 93-day Pancho time series, with a slow rotation and an inclination angle of $60^{\circ}$. The red solid lines correspond to the fitted Gaussian function to the distribution, and the dashed red lines to the mean of the distribution. The black dotted lines correspond to a difference ({\sc fits} -- {\sc inputs}) equal to 0, while the green dashed lines represent the 3$\sigma$ limit ($\sigma$ = standard deviation of the gaussian distribution). The percentages of {\it good fits} within the 3$\sigma$ limit are also given. }
\label{fig:inputfreq}
\end{figure*}

We compared the fitted frequencies with the input frequencies of the simulated data. For both the correct ({\sc a}) and incorrect ({\sc b}) taggings, Fig.~\ref{fig:inputfreq} shows the resulting histograms of the frequency differences between {\sc fits} and {\sc inputs} for the $l = 0$ modes in the case of the 93-day time series (slow rotation and inclination angle of $60^{\circ}$). The percentages of {\it good fits} within 3$\sigma$ are indicated.

\begin{table}
\centering
\caption{Fundamental stellar parameters of Pancho}
\label{table:1}
\begin{tabular}{cccc}\hline
Class & $\log(g) \mathrm{ (cm/s^2)}$ & $\log(Z/X$) & T$\mathrm{_{eff}  (K)}$\\ 
\hline
Dwarf(V) & 4.3 $\pm$ 0.1 & -1.4 $\pm$ 0.1 & 6383 $\pm$ 40 \\
\hline
\end{tabular}
\end{table}

\begin{table}
 \centering
\caption{Global parameters derived from 31-day time series (M$_v$= 9 and 10) of Pancho \citep{Mathur10b}}
\label{table:2}
\begin{tabular}{ccccc}\hline
$\Delta\nu$ & Freq(min) &  Freq(max) & $\nu_{\mathrm{max}}$ & A$_{\mathrm{max}}$\\ 
($\mu$Hz) & ($\mu$Hz) & ($\mu$Hz) & ($\mu$Hz) & (rms ppm) \\
\hline
69$\pm$1.5 &  1000 & 2300 & 1700$\pm$50 & 2.7$\pm$0.3 \\
\hline
\end{tabular}
\end{table}

\begin{figure*}
\centering
\includegraphics[scale=0.62]{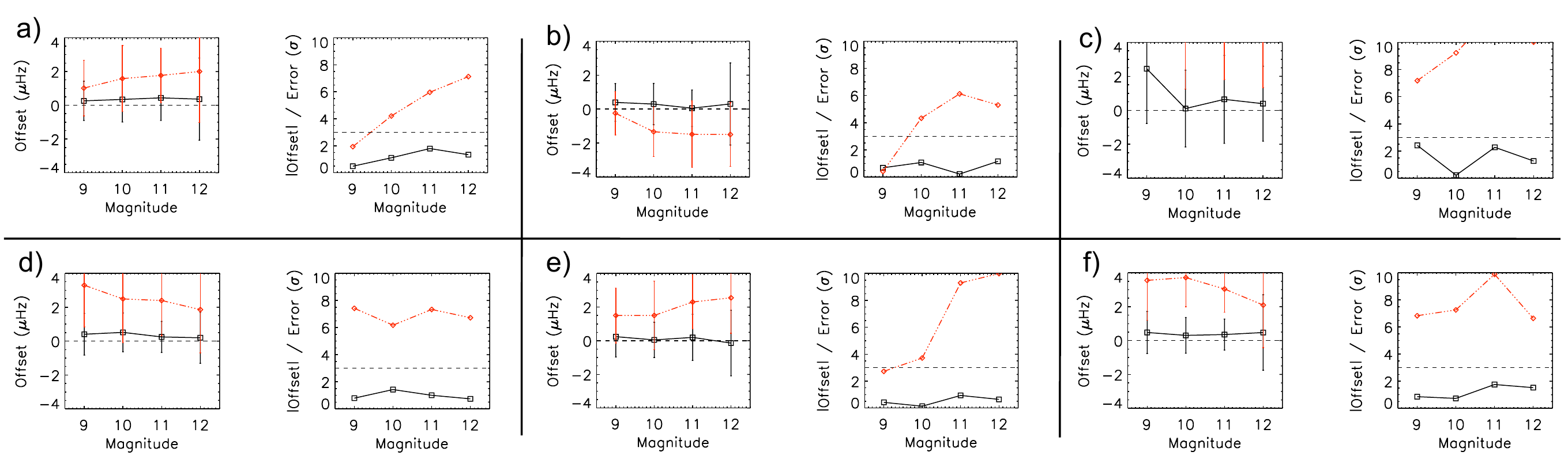}
\caption{Offsets between the fitted and input frequencies in $\mu$Hz (left panels of a), b), c), d), e), and f)) and in sigma unit (right panels of a), b), c), d), e), and f)) as a function of the stellar magnitude M$_v$. The following cases are illustrated: a)  $l= 0$, slow rotation, angle $60^{\circ}$; b) $l=1$, slow rotation, angle $60^{\circ}$; c)  $l=2$, slow rotation, angle $60^{\circ}$; d)  $l=0$, fast rotation, angle $60^{\circ}$, e)  $l=0$, slow rotation, angle $60^{\circ}$, Bayesian conditions; f)  $l=0$, fast rotation, angle $60^{\circ}$, Bayesian conditions. The results for the correct tagging ({\sc a}) and the incorrect tagging ({\sc b}) are represented with black squares and red diamonds respectively.}
\label{fig:inputfreq2}
\end{figure*}

While most of the fits fall within the 3$\sigma$ limit, clear offsets between the fitted and input frequencies are present when the incorrect tagging ({\sc b}) is chosen (Fig.~\ref{fig:inputfreq2}). These offsets of a few $\mu$Hz are larger than 3$\sigma_f$ ($\sigma_f$ = $\langle$formal errors$\rangle$) and tend to increase with stellar magnitude. When the correct tagging ({\sc a}) is chosen, these offsets are minimal and are within the 3$\sigma_f$ limit.

\begin{figure}
\centering
\includegraphics[scale=0.85]{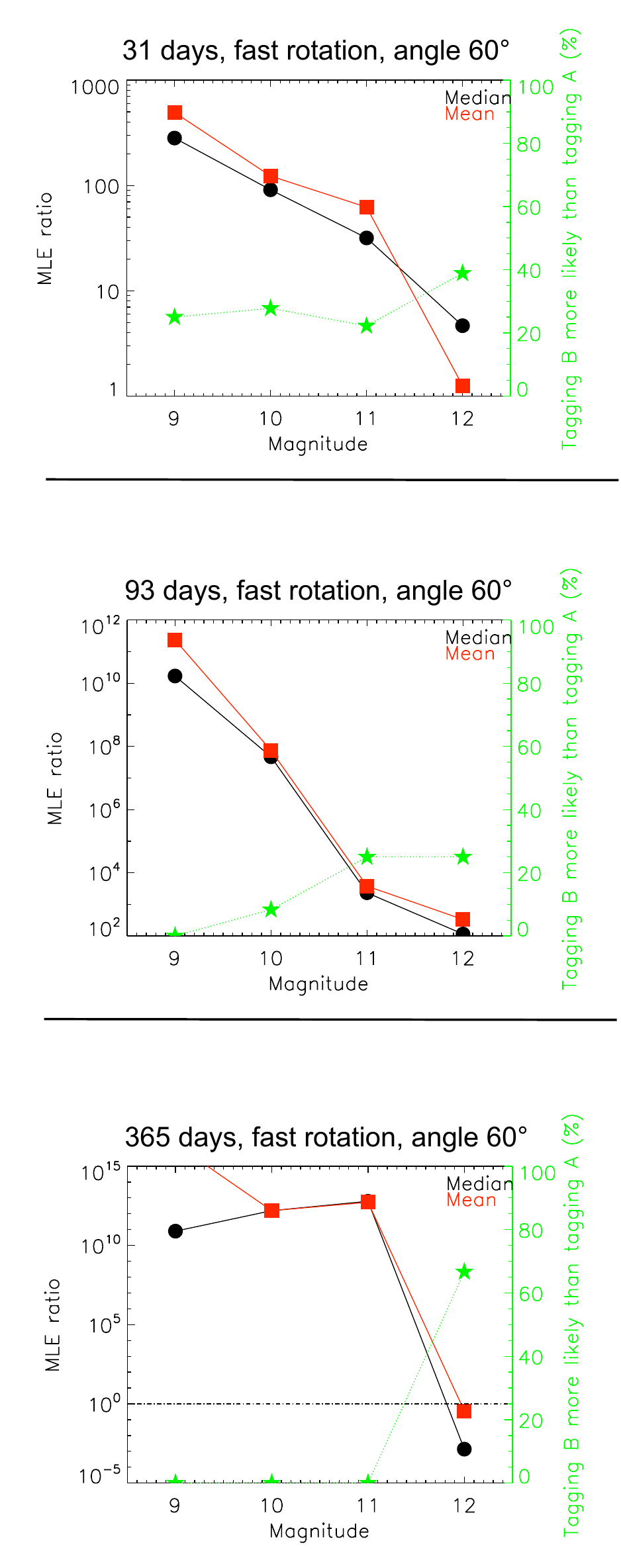}
\caption{Likelihood ratios between the correct ({\sc a}) and incorrect ({\sc b}) mode identifications over the entire set of analyzed power spectra as a function of the star magnitude M$_v$ and the length of observations $T$ (from top to bottom, $T$ = 31, 93, and 365 days respectively) for an inclination angle of $60^{\circ}$. Both mean (red squares) and median (black dots) values of the likelihood ratios are shown. The percentages that the likelihood-ratio test returns the incorrect tagging ({\sc b}) as being more likely are given using the right $y$-axis (green stars).}
\label{fig:mle}
\end{figure}

\section{Likelihood-ratio test}
We present here the reliability of the likelihood-ratio test as a function of the star magnitude M$_v$ and of the length of the observations $T$.
The following discussion concerns the AsteroFLAG star Pancho with fast rotation ($5\mu$Hz) and an inclination angle of $60^{\circ}$.
 For the 3 lengths of observations ($T$ = 31, 93, and 365 days), Fig.~\ref{fig:mle} shows the mean and the median values of the likelihood ratios between the correct ({\sc a}) and incorrect ({\sc b}) mode identifications over the entire set of analyzed power spectra as a function of M$_v$. The percentages that the likelihood-ratio test returns the incorrect tagging ({\sc b}) as being more likely are also represented (right $y$-axis). A clear dependence with M$_v$ and $T$ is identifiable. For the same star and as the length of the time series increases, the likelihood-ratio test is more likely to return the correct answer. However, as the magnitude of the star increases, the likelihood-ratio test is not as reliable.

The percentages that the likelihood-ratio test returns the incorrect tagging (green curves on Fig~\ref{fig:mle}) are:
\begin{itemize}
\item 31-day time series: $\sim$~ 25\% up to M$_v$ = 11;
\item  93-day time series: $\leq$ 25\% up to M$_v$ = 12;
\item  365-day time series: 0\% up to M$_v$ = 11.
\end{itemize}

In the case of stars with slow rotation, these percentages get larger, with for instance, 25\% chance that the likelihood-ratio test returns the incorrect mode identification as being more likely for M$_v$ = 9 and $T$ = 93 days.
The introduction of Bayesian conditions can help mainly when the characteristics of the star make the mode identification difficult, for example, a star with slow rotation.

\section{Conclusions}
The identification of the oscillations modes (or mode tagging) is a necessary step before we can infer the physical properties of the star interiors. Even when using global peak-fitting techniques, the mode tagging remains a difficult task to achieve, when for instance, a large number of oscillation modes are observed with a low signal-to-noise ratio. The likelihood-ratio test can help to discriminate between the possible scenarios (i.e. taggings) and it has been already successfully applied in asteroseismology with the CoRoT, Kepler, and Procyon observations. Nevertheless, by analyzing the AsteroFLAG artificial star Pancho, we showed in this work that the likelihood-ratio test has certain limits depending on both the star magnitude and the length of the observations. For example, for a star with a rotational splitting of $5\mu$Hz and an inclination angle of $60^{\circ}$, observed during 31 days, the likelihood-ratio test will statistically return the incorrect mode identification 25\% of the time for a star magnitude up to M$_v$ = 11. This percentage decreases as the length of observation increases and for time series of 365 days, the likelihood-ratio test will return 100\% of the time the correct identification up to M$_v$=11. However, for stars with slow rotation, the likelihood-ratio test is more {\it likely} to return the incorrect mode tagging. The use of Bayesian conditions help mostly when the star characteristics (for example, slow rotation) make the mode identification difficult.

With the long-term observations that will be collected by the Kepler mission, we hope to have enough signal-to-noise ratio to unambiguously determine the correct identification of the modes for both solar-like stars \citep{Chaplin10} and red giants \citep{Bedding10a}, as well as the stars in open clusters \citep{Stello10}. However the relative faintness of these later stars will probably require the use of the likelihood-ratio test to disentangle between the tagging of the modes.

\acknowledgements   
D.~Salabert acknowledges the support from the Spanish National Research Plan (grant PNAyA2007-62650). This work has been partially supported by the European Helio- and Asteroseismology Network (HELAS) and the CNES/GOLF grant at SAp CEA-Saclay. The authors want to thank the participants of the asteroFLAG meetings at the International Space Science Institute (ISSI) for their useful comments and discussions.


\end{document}